\documentclass[a4paper,12pt]{article}

\usepackage[top=1.2in,right=0.8in,left=.8in,bottom=1.2in]{geometry}

\usepackage{latexsym}
\usepackage{amssymb}
\usepackage{float}
\usepackage{graphics,graphicx}
\usepackage{tabularx}
\usepackage{amsfonts}
\usepackage{amsmath}
\usepackage{enumerate}
\usepackage{booktabs}
\usepackage{multirow}
\usepackage{sectsty}
\usepackage[affil-it,auth-sc]{authblk}
\usepackage{subfig}
\usepackage{color}
\usepackage{mathtools}
\usepackage{mathrsfs}
\usepackage{bmpsize}
\usepackage{xcolor}
\usepackage{siunitx}
\usepackage{hyperref}
\usepackage{xr}
\usepackage{longtable}

\usepackage{multirow}
\usepackage{array}
\usepackage{tabu}
\usepackage{mathtools}
\usepackage{hyperref}
\usepackage{cancel}

\usepackage[normalem]{ulem}               
\newcommand\redout{\bgroup\markoverwith
	{\textcolor{red}{\rule[0.5ex]{2pt}{0.8pt}}}\ULon}
\usepackage{amsmath}
\usepackage{amsfonts}
\usepackage{amssymb}
\usepackage{siunitx} 

\usepackage{enumerate}
\usepackage{sectsty}
\usepackage[affil-it,auth-sc]{authblk}
\usepackage{color}
\sectionfont{\normalsize}
\subsectionfont{\normalsize}

\usepackage{fancyhdr}            
\fancypagestyle{plain}{%
  \fancyhead{}                                     
  \fancyfoot[CE,CO]{\thepage}       
  \fancyhead[CE,CO]{\textcolor[rgb]{0.64,0.15,0.15}{Published in \emph{Proceedings of the Royal Society A} (2019),  \textbf{475}, 20190283 
  \\
doi: \href{http://dx.doi.org/10.1098/rspa.2019.0283}{10.1098/rspa.2019.0283}
}}
\setlength{\headheight}{14.5pt}}

\newcommand{\footmsg}[1]{%
  \let\temp\thempfn%
  \def\thempfs{}
  \footnotetext{#1}
  \let\tempfn\temp}

\begin{document}
	
	\title{Omnidirectional flexural invisibility of multiple interacting voids in vibrating elastic plates}
	
	\author{
		D. Misseroni$^{1}$, A.B. Movchan$^{2}$ and D.  Bigoni$^{1}$\footnote{Corresponding author:\,\, e-mail:\, davide.bigoni@ing.unitn.it; phone:\,+39\,0461\,282507.}}
	
	 \affil[]{$^{1}$DICAM, University of Trento,
	 	via~Mesiano~77, I-38123 Trento, Italy.\\
	 	$^{2}$Department of Mathematical Sciences, University of Liverpool, Liverpool L69 3BX, UK.}
	
	\date{}
	\maketitle

	\begin{abstract}
		In elasticity, the design of a cloaking for an inclusion or a void to leave a vibrational field unperturbed by its presence, so to achieve its invisibility, is a thoroughly analyzed, but still unchallenged, 
		mechanical problem.
		The \lq cloaking transformation' concept, originally developed in electromagnetism and optics, is not directly applicable to elastic waves, 
		displaying a complex vectorial nature. Consequently, all examples of elastic cloaking presented so far involve complex design and thick coating skins. These cloakings often work only for problems of unidirectional propagation, within narrow ranges of frequency, and considering only one colaked object. 
		Here, a new method based on the concept of reinforcement, achieved via elastic stiffening and mass redistribution, is introduced to cloak multiple voids in an elastic plate.  This simple technique produces invisibility of the voids to flexural waves within an extremely broad range of frequencies and thus surpassing in many aspects all existing cloaking techniques.
		The proposed design principle is applicable in mechanical problems ranging from the micro-scale to the scale of civil engineering. For instance, our 
		results show how to design a perforated load-bearing building wall, vibrating during an earthquake exactly as the same wall, but unperforated, a new finding for seismic protection. 
		
	\end{abstract}
	
	Research on metamaterials started from electromagnetism, in an attempt to use subwavelength microstructures to overcome limitations of standard materials 
		and challenge 
	
	properties such as negative refraction, focusing with Veselago lens, topological surface states, and cloaking to render objects (such as inclusions or voids)  invisible~\cite{matlack}.
	The quest for invisibility originated from the first experimental demonstration of microwave cloaking~\cite{schurig}, which has led to significant  developments in the design and the analysis of electromagnetic metamaterials, based on the introduction of the concept of cloaking transformation\cite{Fleury,Cummer2016}.
	
	In elasticity, the invisibility of an inclusion embedded in a material means that the inclusion does not perturb an ambient vibration propagating within the medium.  To reach this objective, the inclusion should be \lq cloaked', namely, surrounded by a coating which produces its invisibility: this is 
	the subject of the present article in the case when the elastic solid is a thin plate subject to flexural waves and the objects to be made invisible are square voids. 
	
	\lq Perfect cloaking' of an object (ideally of {\it any size}) implies {\it omnidirectional} and {\it total scattering
		elimination} over a {\it wide frequency range}, a goal that should be achieved with a sufficiently {\it thin coating}, of {\it simple design and realization}\cite{Fleury}. 
	Would perfect coating be achieved, {\it multiple coated objects should all remain invisible}, a requirement less obvious than it may appear, because multiplicity implies interaction, 
	which may attenuate or even eliminate the coating effect.
	
	The definition of \lq perfect cloaking' can be further specified in terms of the following 7 invisibility requirements: 
	
	(i.) easiness of coating design and implementation technology; 
	
	(ii.) smallness of the ratio between the thickness of the coating skin and the dimensions of the object to be made invisible; 
	
	(iii.) smallness of ratio between the dimensions of the object to be made invisible and of the hosting medium; 
	
	(iv.) overall scattering suppression; 
	
	(v.) broad bandwidth action; 
	
	(vi.) multidirectionality;
	
	(vii.) invisibility of multiple coated objects.

	Note that requirement (iii.) is seldom mentioned in the literature, but it is rather obvious that invisibility of a small object in a large ambient field 
	is easier to be achieved than invisibility in a narrow ambient. In fact, a small object represents a perturbation in an infinite field, so that at a certain distance 
	it spontaneously becomes invisible (in the same vein, a defect in an infinite elastic solid, or a perturbation applied at one end of a long elastic beam, produces a 
	quickly decaying perturbation\cite{gurtin, bertoldi}). 
	The difficulty in the  achievement of the above requirements was clear in the field from the very beginning, so that research focused on possible simplification of the invisibility requirements\cite{Fleury}. 
	The state of the art reported below (with reference to the seven above-listed requirements) shows how the realization of invisibility is still largely unchallenged, particularly in elasticity (a more detailed 
	description can be found in the supporting material).  
	
	(i.) Cloaking has usually\footnote{
		A promising exception to the rule of complication is fiber reinforcement\cite{olsson}, which however works only for special geometrical arrangements.
	} 
	been related to the use of (in the words of the authors of the papers referenced below) \lq exotic materials'. 
	In elasticity: pentamode\cite{Chen2015}, Cosserat\footnote{
		The asymmetry of stress has been advocated as one of the possible means to achieve cloaking in elasticity  \cite{BrunAPL, NorrisWaveMotion}. However, detailed analysis of cloaking effects based on elastic models displaying lack of stress symmetry, such as for instance Cosserat materials, remains a challenge. We also note that  materials with a microstructure possess strain gradient effects, which are \lq higher-order effects' \cite{lakes1, lakes2, bacigallo1, bacigallo2, rizzi1, rizzi2}. However, the question of importance of such \lq effects' in the context of cloaking may require further studies.
	} \cite{BrunAPL, NorrisWaveMotion}, anisotropic density\cite{norris}, prestressed\cite{JonesIntJSolsStruct}, 
	nonlinearly prestrained\cite{NorrisPRS} materials, application of in-plane body forces\cite{BrunNJP, ColquittJMPS}, or of gyroscopic, or chiral elements\cite{Nieves1, Nieves2, Nieves3} have been proposed for the cloaking, all mechanical set-ups posing formidable difficulties of implementation, which have led to 
	extremely complicated cloaking geometries\cite{FarhatSebastienPlates2, Wegener2012, Misseroni_NSR}.

	(ii.) The importance of small thickness for the cloaking skin was pointed out, together with the indication of a route to ultrathin coating, in optics\cite{Xingjie2015}. 
	In elasticity, the ratio between the thickness of the coating skin and the dimension of the coated object
	is sometimes infinite\cite{ParnellPRS, ParnellWaveMotion, NorrisPRS}, it is 3 for a nonlinear flexural cloak\cite{darabi}, 
	it goes down to 0.41 and 0.5 for lattice materials\cite{WegenerPNAS, ColquittPRS} and touches 0.17, but in the quasi-static case\cite{UnfealabilityCloak}. 
	Finally, for flexural waves the thickness ratio is 1.5\cite{Wegener2012} and 0.5\cite{Misseroni_NSR} in the only few experimented cases, and in theoretical treatments  
	5, 1.6\cite{BrunNJP}, 1.17\cite{FarhatSebastienPlates2}, and  0.5\cite{ColquittJMPS, FarhatSebastienPlates, JonesIntJSolsStruct}.

	(iii.) 
	The ratio between the dimensions of the hosting medium and 
	of the coated object (which 
	should be kept as small as possible, to reveal that the coated object is more than a mere perturbation 
	to a large field) 
	is often assumed to be infinite in theoretical considerations\cite{ParnellPRS, ParnellWaveMotion, NorrisPRS, norris, ColquittPRS, BrunNJP, FarhatSebastienPlates, FarhatSebastienPlates2, ColquittJMPS, JonesIntJSolsStruct, WegenerPNAS} 
	and 
	realized as large in the experiments, namely, 
	4.3 for quasi-static cloaking\cite{UnfealabilityCloak}, 
	5 \cite{Misseroni_NSR} and 8 \cite{darabi} for flexural waves, and not reported in another case\cite{Wegener2012}.

	(iv.) Obviously, the chief characteristic of cloaking is overall scattering suppression, a condition which is often reached only in an 
	approximate sense. Quantitative data on scattering reduction 
	are usually not explicitly reported in elasticity\cite{BrunNJP, FarhatSebastienPlates, FarhatSebastienPlates2, ColquittJMPS, JonesIntJSolsStruct, Wegener2012}. 
	A performance of 60\% \cite{Misseroni_NSR} and 55\% \cite{darabi} is reported for flexural waves.

	(v.) Invisibility should be achieved for a wide band of wavelengths, but this is very difficult both in electromagnetism and in acoustics, and in elasticity the situation 
	is worse, due to the vectorial nature of the wave propagation. 
	In the latter field cloaking is usually achieved only at certain frequencies\cite{MiltonNJP, Guenneau, Chen2015, BrunAPL, ColquittJMPS, FarhatSebastienPlates, Guenneau_NSR, ColquittPRS, Misseroni_NSR, FarhatSebastienPlates2,Brule2014,Brule2017,Miniaci2016}. 
	As a matter of fact, data on cloak performances are usually not reported for a continuous spectrum of frequencies (a counterexample to this tendency is \cite{darabi}), so that we have used an available software
	for cloaking in vibrational thin plates\cite{Misseroni_NSR} 
	to check that cloaking obtained through transformation elasticity is working only at certain frequencies.

	(vi.) Ideally, invisibility should concern waves coming from every direction (emanating from a punctual source, reflected from boundaries, interacting with non-invisible objects). In elasticity the source of the waves if often a pulsating force acting in an infinite medium\cite{ParnellPRS, ParnellWaveMotion, NorrisPRS, FarhatSebastienPlates, FarhatSebastienPlates2, ColquittPRS, ColquittJMPS}. Some cases are restricted to unidirectionality\cite{BrunNJP, Misseroni_NSR}.

	(vii.) Only a limited theoretical effort\cite{far2}, without experimental work, has been so far reported to render multiple objects invisible. This \lq multiple invisibility' is 
	particularly important when the objects are located at a small relative distance, so that a strong interaction between them would occur when cloaking is absent.

	It can be concluded that in planar elasticity even imperfect and narrow band cloaking 
	is still far from being achieved and no experiments of any kind have been conducted so far, with the exception of the so-called \lq quasi-static cloaking'\cite{UnfealabilityCloak}, which was obtained only in an imperfect sense and under the restriction of unidirectional load (it was also investigated for lattice materials\cite{WegenerPNAS}). The quasi-static cloaking is a well-known concept in mechanics\cite{bertoldi, ru, milton}, but is 
	far from satisfying the strong requirements needed to produce invisibility when dynamic vibrations are considered.

	The only experiments (three papers\cite{Wegener2012,  Misseroni_NSR, darabi}) available for cloaking in elasticity refer to plates subject to flexural vibrations, to which the present article is addressed. The experiments report (i.) extremely complicated (20 concentric rings and 16 different elastic materials in one case\cite{Wegener2012}, a cloaking designed on a convoluted geometry obtained from transformation elasticity in \cite{Misseroni_NSR}, and 15 and 30 layers in a nonlinear design \cite{darabi}),
	(ii.) thick cloaks (thickness ratios 1.5\cite{Wegener2012}, 0.5\cite{Misseroni_NSR}, and 3\cite{darabi}), (iii.) operating on small objects (smallness ratio 5\cite{Misseroni_NSR}, 8 \cite{darabi}  
	and not reported\cite{Wegener2012}), (iv.) with low scattering suppression ($\approx$~60\%\cite{Misseroni_NSR}, not reported\cite{Wegener2012}, and 55\%\cite{darabi}), (v.) restricted to a narrow bandwidth (the cloak works only at certain frequencies within a 200 Hz range\cite{Misseroni_NSR, Wegener2012}, a broadband is reported in \cite{darabi}, but only in the high frequency range of 2-11 kHz, (vi.) unidirectional\cite{Misseroni_NSR}, finally, (vii.) 
	invisibility of multiple objects has never been attempted.

	The problem of an elastic plate is addressed here, containing a series of square voids and subject to flexural, time-harmonic vibrations. 
	A radically new design concept is introduced to make voids invisible within an extremely broad range of frequencies, never challenged so far. 
	The cloaking design is first introduced with reference to structured plates (a rectangular lattice of flexural and torsional elastic rods) with one void and its unprecedented capabilities are demonstrated with a 
	numerical analysis of the coefficients in a multipole expansion of the scattered field. Subsequently, the cloaking is applied to an elastic plate perforated with three square voids positioned at a close distance to each other and even in this more complex case the system is shown (both experimentally, with a new experimental set-up, and numerically) to behave as an intact plate, up to the fourth flexural mode of vibration and beyond, so that it can be concluded that the two systems are equivalent for dynamic flexural vibrations. 
	Finally, a series of three and of seven voids are considered with random distribution and inclination with respect to the wave propagation direction, so that here {\it torsion} is excited and coupled with flexion. Even in this case, the cloaked voids display only a small difference in the vibrational modes when compared with those relative to the intact plate, for frequencies up to 400 Hz and beyond. 
	
	We suggest that the performance of a cloaking should always be referred to the above-listed 7 requirements\footnote{
		Imagine for instance that one drills a little hole in a wall internal to a big skyscraper. It would be not surprising that the little hole 
		does not influence the seismic behavior of the huge building, so that the little hole would be invisible, but in this case the requirement (iii.) helps  
		in evaluating the situation. Another example will be presented later when it will be shown that a bad cloaking can exceptionally work very well at certain frequencies, 
		a situation which clearly violates the requirement (v.). These simple examples show why a cloaking system should always be presented with reference to the 7 listed requirements.
	}
	, so that we point out that 
	the cloaking technique proposed in the present article introduces:

	(i.) an extremely simple design (based on a stiffness and mass redistribution) and realization (only an elastic stiffening frame with punctual masses surrounds the voids to be cloaked) of the cloak, 
	
	(ii.) the smallest (5.997~mm/45~mm~$\approx 0.13$) value of the ratio between thickness of coating skin and dimensions of the object to be made invisible ever attained for elasticity, 
	
	(iii.) 
	a small ($\approx$~0.58) value of ratio between the 
	dimensions of the object to be made invisible and the 
	dimensions of the hosting medium (in the cases treated in this paper the voids introduce a porosity ranging between 4\% and 15\%, never challenged before), 
	
	(iv.)  an excellent scattering suppression (which goes up to 99.75\% for certain frequencies and geometries), 
	
	(v.) the broadest invisibility bandwidth ever reached (from 0 to 400 Hz and beyond, for instance, the cloaking for flexural vibration of aligned voids was found to work well from 0 up to 750 Hz), 
	
	(vi.)  an excellent multidirectionality, never explicitly shown before and demonstrated allowing square voids to be inclined with respect to the propagation of the waves, 
	
	(vii.) the demonstration of invisibility for multiple coated objects (randomly distributed and inclined), never attempted so far, allowing interaction between several voids in the uncloaked case.

	\section{Structured plates and the cloaking design}

\begin{figure}[!h]
		\renewcommand{\figurename}{\footnotesize{Fig.}}
		\begin{center}
			\includegraphics[width= 1\textwidth]{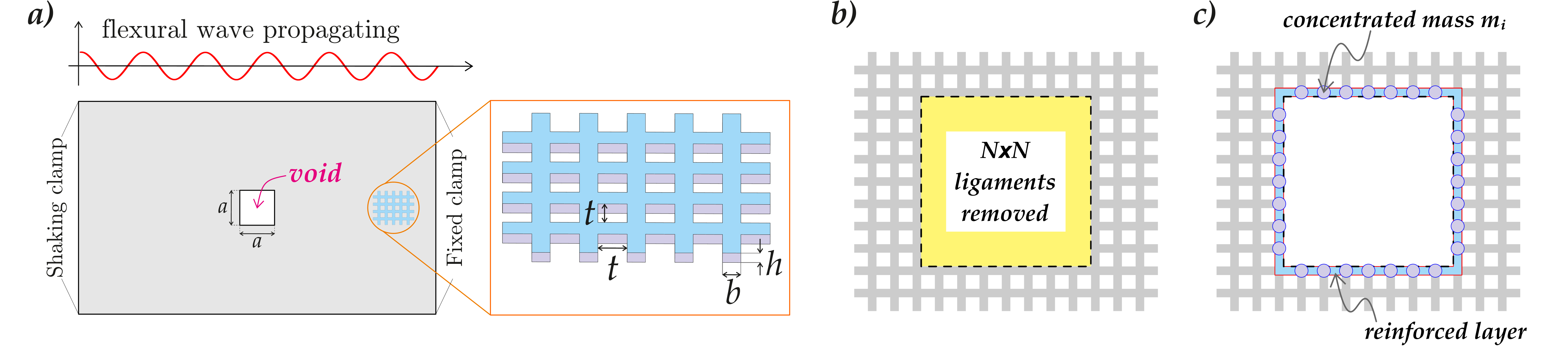}           
			\includegraphics[width= 1\textwidth]{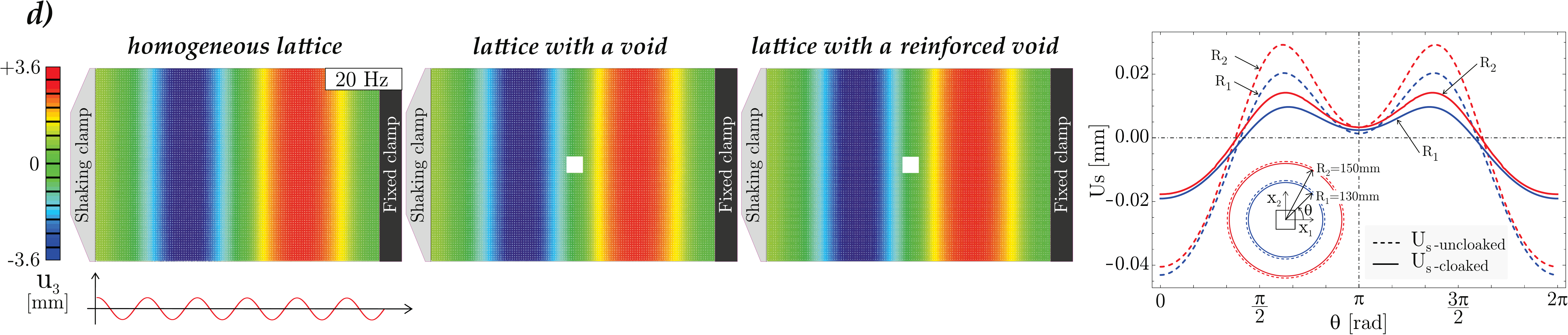}
			\includegraphics[width= 1\textwidth]{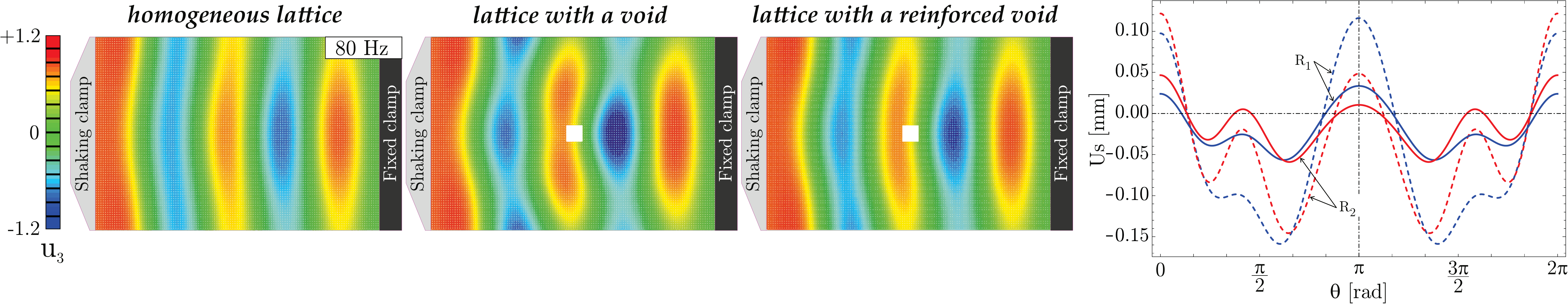}
			\includegraphics[width= 1\textwidth]{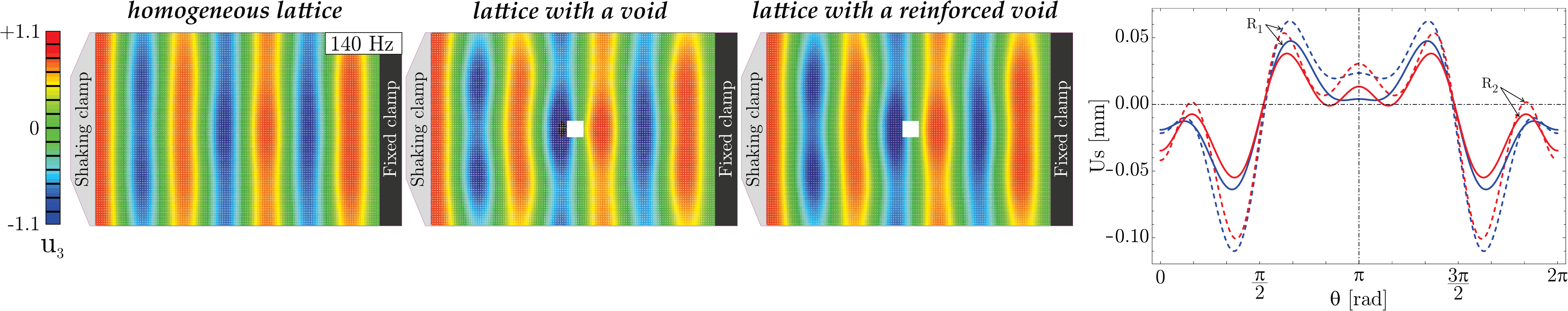}
			\includegraphics[width= 1\textwidth]{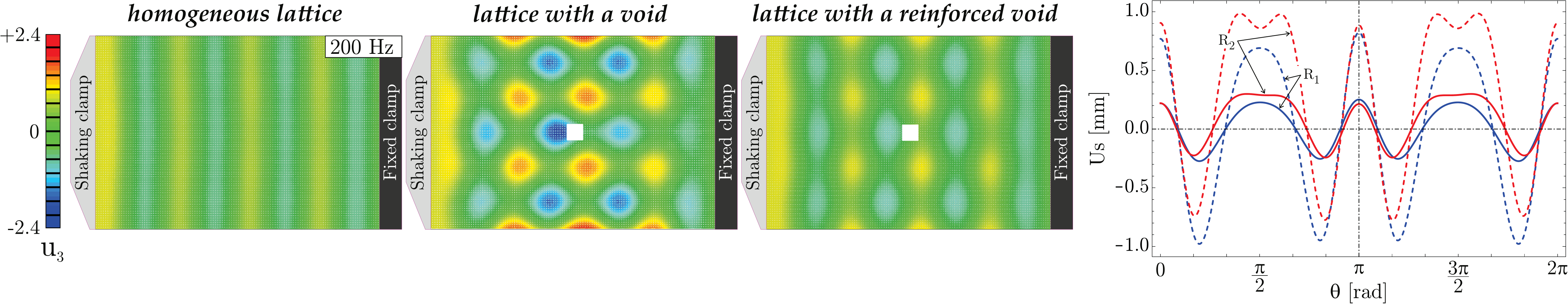}
			\caption{\footnotesize (a) Flexural lattice with a square void, the lattice is detailed in the inset. 
				Note that a sinusoidal displacement with null rotation is imposed on the left end of the plate, while the right end is clamped and the two horizontal ends are 
				left 
				traction-free. 
				(b) Lattice with a square void and (c) reinforcement and  punctual mass redistribution providing the cloaking. 
				(d)  Flexural displacement distribution ($u_3$, [mm]) and scattered field $U_s$ at low and high frequencies evaluated with the finite element program Abaqus, (from left to right) for the uniform structured plate, for the
				structured plate with a square void, and with cloaked void. 
				The scattered field, evaluated along two concentric circles (blue for radius 130 mm and red for 150 mm) for the uniform lattice and 
				for the lattice containing the cloaked square void, is reported on the right.
			} \label{Fig00}
		\end{center}
	\end{figure}

	A structured plate in the form of a rectangular flexural lattice is subject to a sinusoidal incident wave 
	(generated by imposing a sinusoidal displacement with null rotation on one end of the lattice, while the opposite end is clamped and the two remaining ends are left traction-free, Fig. \ref{Fig00}a), and contains a square void, with homogeneous Neumann boundary conditions (in other words, traction free edges, Fig. \ref{Fig00}b). 
	Without recurring to the concept of cloaking transformation, a simple procedure to strongly reduce scattering is proposed,
	based on mass and moment of inertia conservation and elastic stiffening averaging along the boundary of the void. 
	In particular, a (discrete, Fig. \ref{Fig00}c, or a continuous, the latter treated in the supplementary material) 
	redistribution of mass in the lattice is designed around the boundary of the void,  
	maintaining the same mass eliminated by the void and the same moment of inertia around the axes of the void.
	Moreover, the elastic ligaments surrounding the void (Fig. \ref{Fig00}c) are stiffened 
	in such a way to maintain the same bending and torsional stiffness. This can be achieved in different ways, one is to properly change the Young modulus of the elastic ligaments.
	Adopting this cloaking design, numerical simulations show a drastic reduction in the coefficients of the multipole expansion of the scattered field for 
	the reinforced void. 
	
	The proposed procedure can be developed further to a full-scale optimal design algorithm, but our objective now is to keep the approach simple and efficient, for frequencies up to wavelengths comparable to the size of the void in the lattice. The introduced cloaking technique is proven to yield 
	scattering reduction for a range of frequencies much broader than those achieved so far with different approaches.
	
	The flexural lattice with the square void is shown in Fig. \ref{Fig00}.  
	Part (a) of the figure shows the intact lattice, whereas the lattice with the uncloaked void is sketched in part (b), while the cloaking design is shown in part (c), where the boundary of the void has been reinforced with a stiffening frame and a discrete distribution of masses.
	Furthermore, two circles enclosing the void are indicated in the inset of Fig. \ref{Fig00}d, along which the Fourier coefficients have been evaluated for the scattered field.
	
	In the structured plate, each ligament of the flexural lattice obeys, for time-harmonic out-of-plane displacement ${ u}$ and torsional rotation $\theta$, 
	both of pulsation $\omega$, 
	the equations of motion, 
	$$
	D \frac{d^4 u(x)}{d x^4} - \rho bh\,\omega^2 u(x) =0,  ~~~ \frac{d^2 \theta(x)}{d x^2} + \omega^2 \frac{\rho I_p}{C} \theta(x) = 0,
	$$
	where $x$ is a local variable running along one of the thin ligaments (having mass density $\rho$, rectangular $b \times h$ cross-section, 
	polar moment of inertia $I_p$, torsional rigidity $C$, and flexural rigidity $D=b\, h^3 E/12$, where $E$ is the Young modulus) of the lattice.  The following values were assumed for the lattice ligaments: $E$=3350~MPa, $\nu=0.3$, 
	$\rho=1200$~kg/m$^3$, $b$=3~mm, $h$=2~mm, so that a structured plate of dimensions $L$=605~mm, $W$=405~mm has been considered. 
	At the junction points between the thin ligaments, the flexural displacement and its first derivatives are continuous, and, in addition, two balance conditions for the total bending moment and the total transverse force are satisfied. The two above differential equations must obey the junction conditions at the nodes of the grid, 
	imposing continuity of displacement and rotation. 
	In the simulations below, the structured plate is subject to a prescribed sinusoidal displacement with null rotation at one edge and clamped at the other, while the other two edges are traction free.
	
	Three different geometrical settings are used: (i.) 
	the perfectly uniform flexural lattice (without any void);
	(ii.) the lattice with a square void without reinforcement and without any alteration of mass along its boundary;
	(iii.) the lattice with the square void already considered, but now with the boundary both reinforced and with altered mass, as described below.
	A simple design is outlined for the inertial reinforcing layer around a void in a square flexural lattice, showing that scattering due to the interaction of the incident wave is reduced down to negligible values, within certain frequency intervals.
	This design is different from earlier work \cite{Misseroni_NSR} and has the advantage of simplicity of implementation, as well efficiency of the scattering reduction.
	Assuming an array of $N \times N$ junction points inside the square region, corresponding to the average mass $M$ assigned to each junction, the total mass of ${\cal M}=N^2 M$ has to be \lq redistributed' along the boundary layer surrounding the void. 
	The additional mass is added to each node (except to the corners) on the interior reinforcing layer and is denoted by $m$. 
	At the corner point of the square void, no mass is added, so that $4N m = N^2 M$, or equivalently
	$m = {\cal M}/(4 N)$. Note that for simplicity the mass has been redistributed only to match the total mass of the lattice inside the square region, but not
	its moment of inertia, which should be matched for a more refined design of the mass re-distribution. 
	Additional reinforcement is imposed through stiffening of the tangential ligaments running along the boundary of the square void as follows.
	The elastic Young modulus of the tangential ligaments is selected in a way so that both the flexural and the torsional stiffness 
	of the \lq missing ligaments' is added to their own stiffness. The additional stiffness can be easily calculated, as the stiffnesses of several ligaments simply sum up to provide the 
	equivalent stiffness.
	The Young modulus of the tangential ligaments results to be equal to 15075 MPa.
	
	\subsection{Scattering reduction}
	
	Finite element simulations (using Abaqus) have been performed to demonstrate the efficacy of the reinforcement, so that the results presented here deliver a comparison between the waves in the homogeneous intact lattice, a lattice with a square void, and a lattice with the reinforced void. The simulations were performed for a selected range of frequencies, with the emphasis on  the scattering produced by a void with traction free boundary. 
	The steady-state frequency response of each lattice was computed using the dynamic/explicit package	implemented in Abaqus. The simulations, performed using a parametric python script, were conducted using 3D beam elements.
	In addition to the visual assessment of the front of the flexural wave, measurements of the scattered displacement field are performed along the circular paths shown in the inset on the right of Fig.\ref{Fig00}d. The radii of the circles (130 mm and 150 mm) are much larger than the size of the void, and the centers of the circles are positioned at the center of the void.
	The scattered displacement field is defined as the difference between the displacement in the flexural plate with the (uncloaked and cloaked) void and the displacement field in the homogeneous flexural plate without voids.
	A selection of computational results is reported in Fig. \ref{Fig00}d (more information is provided in the supplementary material), for frequencies ranging between 20 Hz and 200 Hz.
	The diagrams in this figure include the color maps of the flexural displacement for: (i.) the intact lattice (without void);
	(ii.) the lattice with the square void (without reinforcement and mass redistribution); (iii.) 
	the lattice with the square void cloaked through a stiffening boundary layer where the mass is redistributed.
	In the diagrams showing the amplitudes of the scattered fields, the solid lines represent the scattered flexural displacement around the reinforced void, whereas the dashed lines correspond to the scattered flexural displacement around the square void without reinforcement and mass redistribution. A comparison has to be made between the data evaluated on the circle of the same radius, in particular, data have been reported with a blue (red) line for the circle of radius 130 mm (150 mm).
	It can be observed from Fig. \ref{Fig00}d, upper part, that at low frequency (20 Hz) the presence of the void, reinforced or not, does not alter much the response. However,
	the cloaking effect of the reinforcing layer becomes clearly evident starting from the frequency of 40 Hz and up to 200 Hz (see also the supporting material). 
	The wave front in the homogeneous plate is planar and its distortion produced by the presence of the unreinforced void becomes distinctly visible at 80 Hz, but this distortion is strongly reduced in the case of the cloaked void. 
	At high frequencies, one would expect that the procedure based on the reinforcement of the boundary and redistribution of the inertia could become less accurate. Nevertheless,  a significant reduction of scattering is observed for cloaked voids, clearly appreciable when the fields are compared to those pertinent to the void without reinforcement and mass redistribution. The beneficial effect of the reinforcement is observed up to 200 Hz and beyond (data not reported for brevity).

	\subsection{Evaluation of the multipole coefficients for the scattered field}
	
	Quantitative data on scattering reduction can be obtained for reinforced void, by considering the scattered flexural displacement field $u_s$, represented along a circle of  radius $R$, concentric to the square void, in the form
	\begin{equation}
	\label{fessa}
	u_s(R, \theta) = \sum_{i=0}^N C_i(R) \cos(i \theta), 
	\end{equation}
	where $\theta$ stands for the polar angle, $\theta \in [-\pi,  \pi),$ and $R$ is the radius.
	The geometry of the elastic solid and the applied displacement are symmetric with respect to the horizontal axis, so that the series
	(\ref{fessa}) represents an even function of $\theta$.

	The moduli of the Fourier coefficients $|C_i(R_1)|, |C_i(R_2)|, i=0,...,4,$ were evaluated on the data obtained from the numerical simulations for the scattered flexural displacement along the circular paths of radii $R_1$=130 mm and $R_2$=150 mm, Fig. \ref{Fig3a} (more information is provided in the supplementary material).  In these diagrams, circular spots are used to mark values corresponding to the Fourier coefficients of the scattered displacement around the square void without reinforcement and mass 
	redistribution, while the triangular markers are used for the \lq cloaked' square void (with the reinforced boundary and redistributed mass). Fig.  \ref{Fig3a} covers frequencies ranging between 20 and 200 Hz. The index $i$ used in the figures denotes the index of the coefficient in the series (\ref{fessa}). The triangular markers always lie below the circular ones, so that results clearly indicate that the computed Fourier coefficients for the cloaked void have moduli remarkably smaller than those pertaining to the uncloaked void.
	
	A direct comparison between different cases is straightforward from Fig. \ref{Fig3a}, so that the orders of magnitude of $C_i(R)$ for given 
	frequencies can be appreciated. For example, at 100 Hz  the maximum value of coefficients is smaller, approximately by 10 times, than the maximum value at the high  frequency of 200 Hz. The computations confirm the efficiency of the reinforcing cloak accompanied by the redistribution of mass.

	\begin{figure}[ht]
		\renewcommand{\figurename}{\footnotesize{Fig.}}
		\begin{center}
			\includegraphics[width= 1\textwidth]{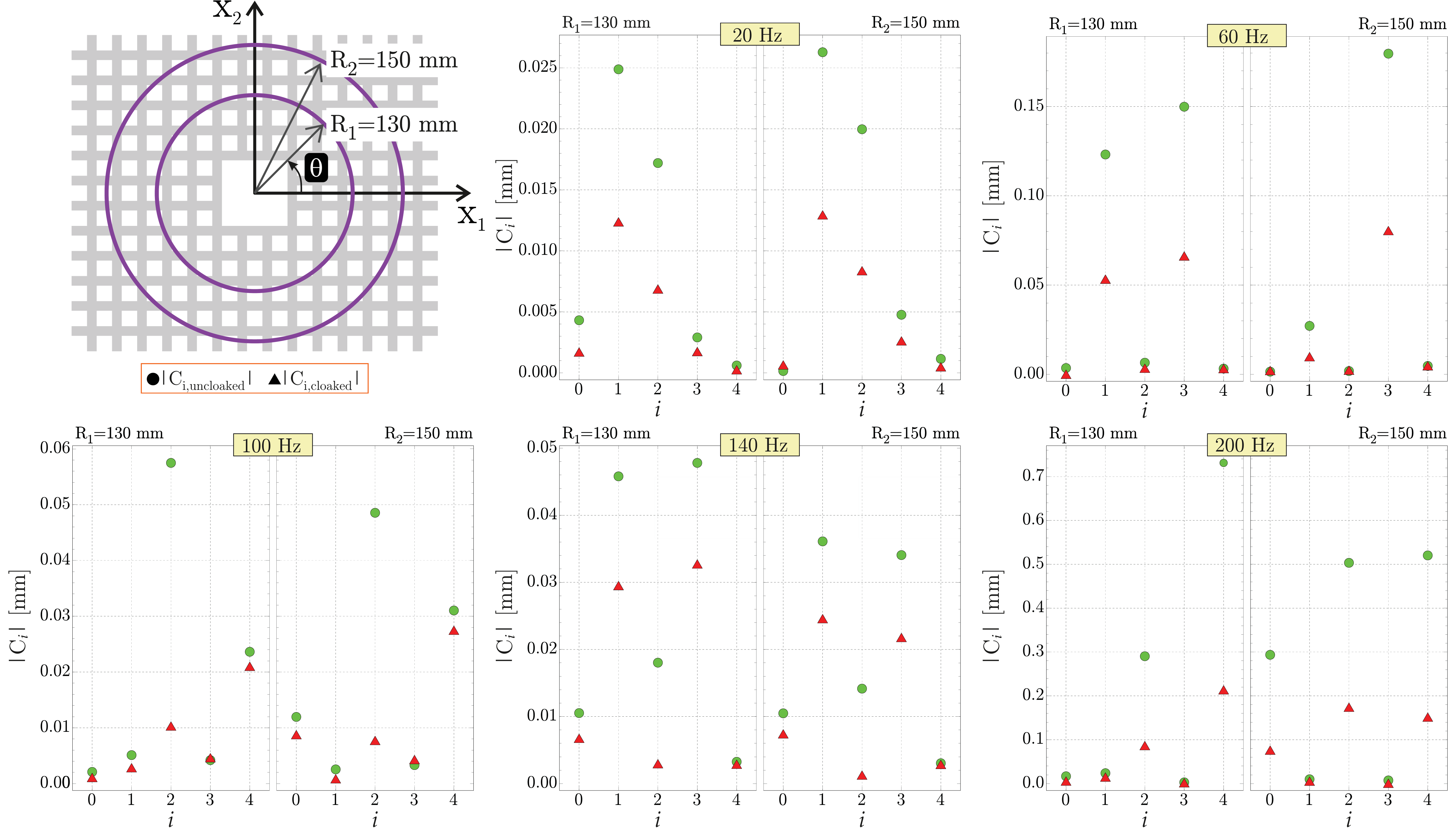}
			\caption{\footnotesize 
				Fourier coefficients for the representation of the scattered displacement around a square void in a structured plate subject to sinusoidal vibration, note that index $i$ is pertinent to the series (\ref{fessa}). Frequencies range between 20 and 200 Hz and coefficients are evaluated with reference to the two circular contours of the radii $R_1=$130 mm and $R_2=$150 mm. Results show that the moduli of the Fourier coefficients for the case of the cloaked void are smaller than those referring to the case of uncloaked void.
			} \label{Fig3a}
		\end{center}
	\end{figure}

	\section{Elastic plates}
	
	The proposed reinforcement design for structured plates can be readily applied to an elastic perforated plate subject to flexural and torsional vibrations, to render the voids invisible.
	\begin{figure}[h]
		\renewcommand{\figurename}{\footnotesize{Fig.}}
		\begin{center}
			\includegraphics[width= 1\textwidth]{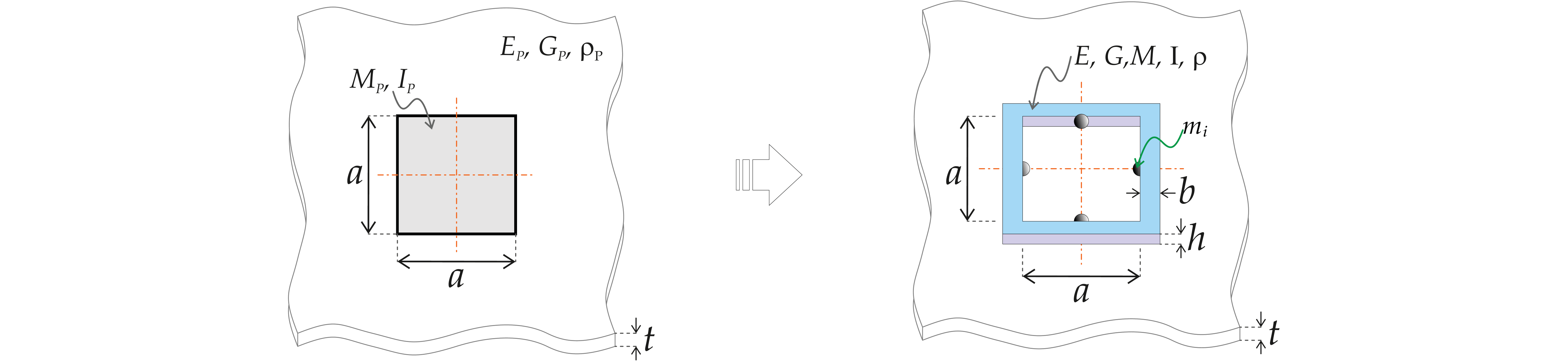}
			\caption{\footnotesize 
				A sketch of the cloaking design. A square $a \times a$ void has to be cloaked in an elastic plate (left), so that a square void of dimension $(a+2b)\times (a+2b)$ is cut off from the 
				plate, and a reinforcement frame of thichkess $b$ is attached, which provides the cloaking (right). The reinforcement frame has bending and torsional stiffnesses and mass and moment of inertia equivalent to the material eliminated with the cut. 
			} 
			\label{FigPOC_design}
		\end{center}
	\end{figure}

	The design of the cloaking is performed into different steps (Fig. \ref{FigPOC_design}). 
	\begin{itemize}
		\item A void of dimension $a \times a$ has to be cloaked in an elastic plate, so that a material square element of dimensions $(a+2b) \times (a+2b)$ (where $b$ is the width of the reinforcement frame) is 
		cut off from the plate. 
		\item An elastic square grid, made up of beams with flexural rigidity $EJ_x=EJ_y$ and bending rigidity $GJ_t$ and spaced at a distance $a/2$, behaves as an isotropic elastic plate of thickness $t$ and Young modulus and Poisson's ratio 
		respectively 
		$E_p$ and $\nu_p$, when\cite{timo}
		\begin{equation}
		\label{match}
		EJ_x = G J_t, ~~~~~~
		\frac{EJ_x}{a/2+b} = \frac{E_pt^3}{12(1-\nu_p^2)}.
		\end{equation}
		\item The flexural and torsional stiffnesses calculated from Equation (\ref{match}) for the grid are used to reinforce the void in the plate. 
		The conditions (\ref{match}) can be enforced with a series of different means. A frame of beams with rectangular $b\times t$ cross section, in which $b>t$, is used here. 
		\item a match of the mass and its second moment of inertia between the material eliminated from the plate and the added reinforcement frame implies
		\begin{equation}
		\label{zozza}
		M_P=M+\sum_{1}^{N}m_i, \quad \quad I_P=I+\sum_{1}^{N}m_i d_i^2,
		\end{equation}
		where $M_p$ is the mass removed from the plate to create the void, $M$ is the mass of the frame reinforcement and $m_i$ are the concentrated masses applied at the reinforcement at distance $d_i$ from the mid axes.
	\end{itemize}

\begin{figure}[!h]
		\renewcommand{\figurename}{\footnotesize{Fig.}}
		\begin{center}
			\includegraphics[width= 1\textwidth]{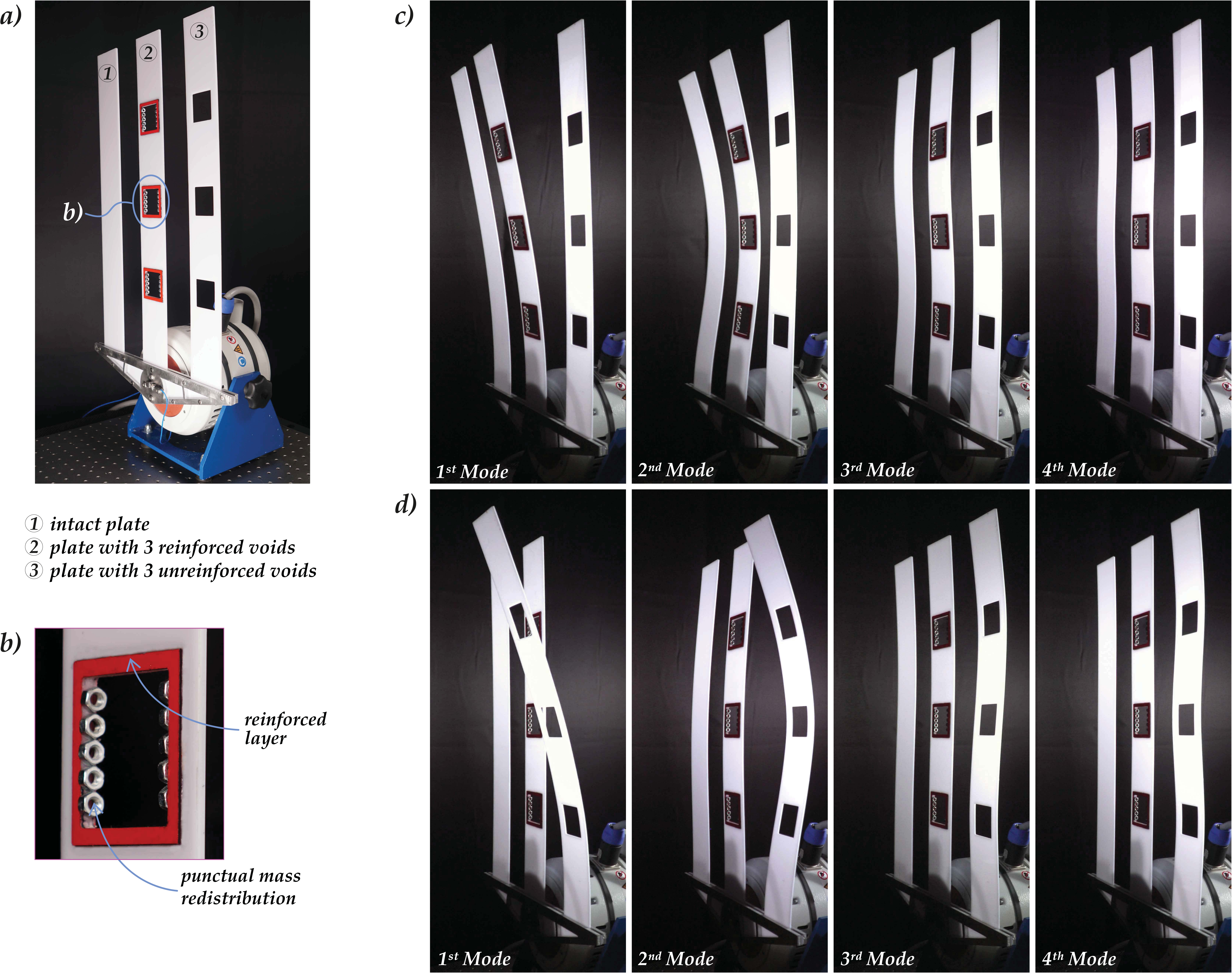}
			\caption{\footnotesize 
				(a) Experimental set-up for the proof-of-concept cloaking model. A shaker imposes a sinusoidal displacement with null rotation at the bases of three elastic plates: one is intact and the other two perforated with three square voids. One of the two perforated plates has the voids cloaked with the proposed reinforcement and punctual mass re-distribution; (b) A detail of the cloaking system; note that the punctual mass redistribution does not affect the stiffness. 
				Parts (c) and (d) show the experimental proof of the cloaking concept. From left to right of part (c): the first 4 vibrational eigenmodes are identical in the intact plate and in  the plate with cloaked voids. Note that the plate with unreinforced voids remains almost undeformed because it vibrates at a frequency far from its resonance. From left to right of part (d): the first 4 vibrational eigenmodes of the plate with uncloaked voids are different from those of the intact plate and of the plate with cloaked voids, which remain almost undeformed, because they vibrate now at a frequency far from their resonance.
			}
			\label{FigPOC_EXP}
		\end{center}
	\end{figure}

	The cloaking is applied to a plate with three square voids of large dimensions, located at a close distance to each other, as shown in Fig. \ref{FigPOC_EXP}, where a proof-of-concept experiment is shown. 
	Note that the boundary conditions are now different from those shown in Fig. \ref{Fig00}, so that now a sinusoidal displacement with null rotation is applied on the lower end of the 
	plate, while all the other ends are left traction-free, Fig. \ref{FigPOC_SIM_quantitative}. These boundary conditions will be realized in the experiments and employed for the rest of the article.
	In the experiments, an electromagnetic shaker (see the supporting material for details) imposes a sinusoidal displacement with null rotation of given frequency at the base of three elastic plates (made up of polycarbonate 3 mm $\times$ 70 mm $\times$ 580 mm), one intact, another perforated with three square voids (each of dimension 45 mm $\times$ 45 mm) and another 
	with the same voids, but cloaked.  
	
	The frequency of the excitation is continuously varied until resonance, to show that the eigenmodes of the intact plate and of the plate with cloaked voids are the same, while the resonance modes of the perforated plate with uncloaked voids are different. 
	This coincidence of resonance frequencies for the intact plate and the plate with cloaked voids determines the invisibility of the voids, 
	because two structures become dynamically identical if {\it all} their resonance frequencies coincide.
	Flexural waves propagate in a plate of stiffness $D$ at frequency $f$ and at velocity $c_f =(D/(\rho A))^{1/4}\sqrt{2\pi f}$, where $\rho$ is the volumetric mass density and $A$ the area of the cross section 
	of the plate. The wavelength and the wavenumber are defined as $\lambda = c_f/f$ and $\kappa = 2\pi/\lambda$.

	The reinforcement was designed in the above-described way, with the following slight adjustment, because only flexural (not torsional) vibrations are analyzed. The same mass of the material eliminated to create the void is redistributed along the two vertical edges only of the void, so to maintain the same inertia moment about the central axis of the void. 
	Moreover, the void (of dimension $a \times a$)  is reinforced to produce the same flexural stiffness of the eliminated material.
	
	The resonance is detected using a stroboscopic light and the experiments exactly confirm the expectation (a set of photos is shown in Fig. \ref{FigPOC_EXP}) and show 
	coincidence up to the fourth mode of vibration for the intact plate and the perforated plate with cloaked voids, while the plate with the unreinforced voids 
	vibrates far from its resonance frequency and thus appears only slightly deformed in the upper part of the figure. 
	The resonance of the latter plate is explored (first 4 modes) in the lower part of the figure, showing that the other two plates behave in the same manner and are now 
	far from resonance. 
	
	Modes higher than the fourth become of difficult detection with the developed experimental set-up, so that 
	the experiments have been complemented with numerical simulations performed with Abaqus showing that the intact plate and the plate
	with cloaked voids display the same dynamical behaviour 
	so that invisibility is achieved.
	(see the supplementary material).

	\section{Omnidirectional cloak and multiple invisibility of voids}
	
\begin{figure}[H]
		\renewcommand{\figurename}{\footnotesize{Fig.}}
		\begin{center}
			\includegraphics[width= 1\textwidth]{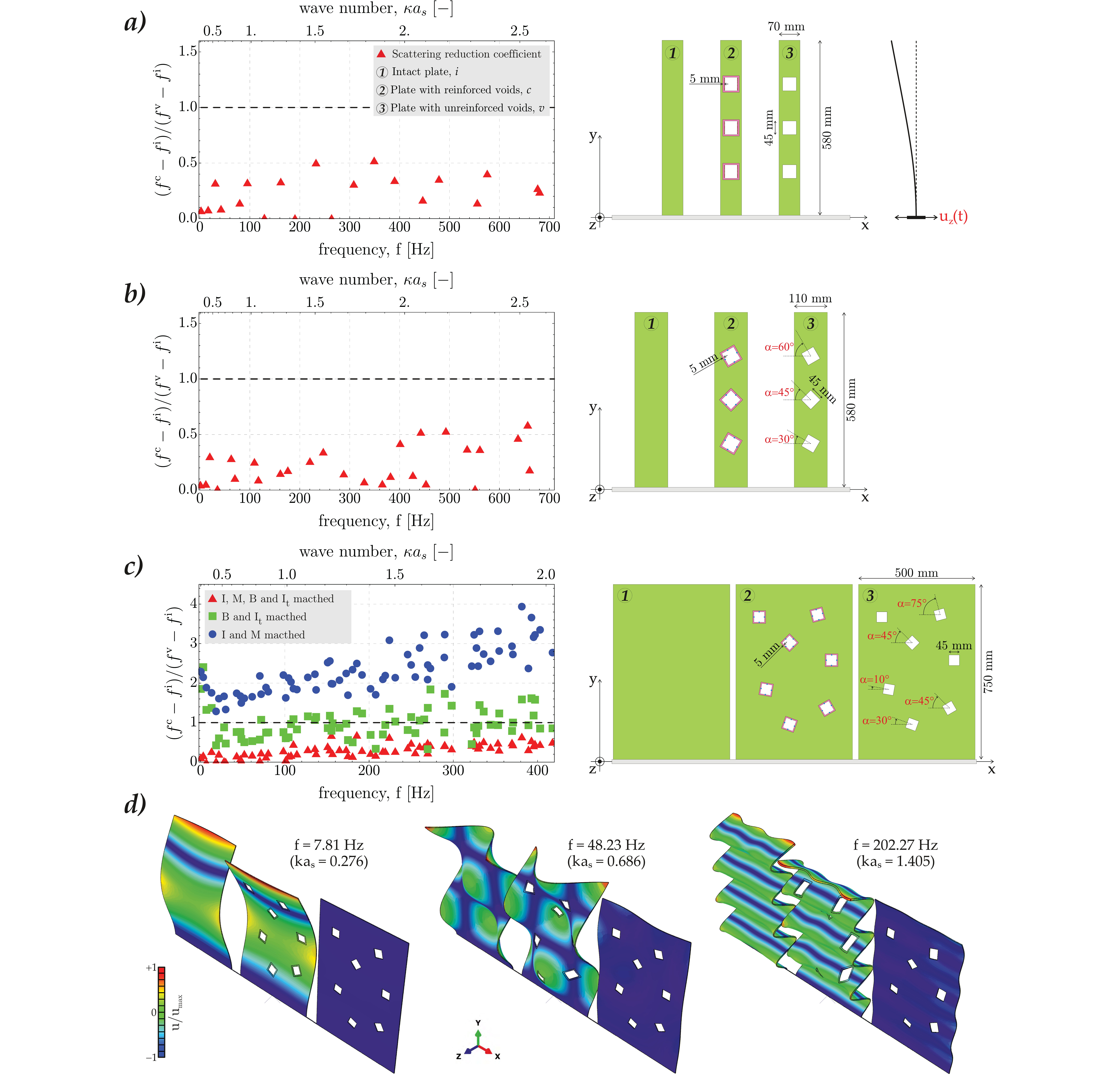}
			\caption{\footnotesize 
				Results from f.e.m. numerical simulations demonstrating omnidirectionality of the cloak and multiple invisibility for several voids randomly located and inclined: 
				scattering reduction coefficient, equation (\ref{ciccio}), 
				is reported as a function of the frequency $f$ on the left 
				for three parallel (a), three randomly oriented (b), and seven randomly located and oriented (c) voids (the set-ups used for the simulations are sketched on the right for the 
				intact plate and the plates with cloaked and uncloaked voids; in all cases a sinusoidal displacement with null rotation is imposed on the lower end of the plate, while all the other ends are traction-free). In part (c), results obtained with a cloaking which matches both stiffness and mass are reported with red markers, while results 
				indicated with green (blue) markers have been obtained by neglecting mass redistribution (stiffness reinforcement).
				Part (d): three eigenmodes at different frequencies, involving 
				coupling between flexural and torsional vibrations, are shown to coincide for the intact plate (left) and the plate with cloaked voids (centre), so that invisibility is 
				verified, 
				and both differ from those pertinent to the plate (right, remaining almost undeformed) with unreinforced voids.
			} 
			\label{FigPOC_SIM_quantitative}
		\end{center}
	\end{figure}

	The experiments and the numerical simulations so far performed refer to cases in which the wave propagation direction is orthogonal to the 
	voids and the plate is subject to flexural vibrations. 
	Omnidirectionality of the proposed cloak is demonstrated through numerical simulations considering
	now the situations in which the voids are randomly inclined  
	with respect to the wave propagation direction and in which seven randomly distributed and oriented voids are  present. 
	In these cases, there is a strong coupling between flexural and torsional modes. 
	The scattering reduction is quantified through the \lq scattering reduction coefficient'\cite{sanchis}   
	\begin{equation}
	\label{ciccio}
	(f^c-f^i)/(f^v-f^i),
	\end{equation} 
	namely, the quotient of the eigenfrequencies relative to the cases with cloaked, $f^c$, and uncloaked, $f^v$, voids, both subtracted from the eigenfrequency relative 
	to the intact plate, $f^i$. Therefore, a perfect cloak realizes a coefficient equal 0, while 1 corresponds to a completely inefficient one and values superior to 1 mean that the
	cloak is worse than nothing (a situation which is often found in the literature\cite{darabi, sanchis}). 
	The reinforcement is obtained now by matching the mass and its second moment of inertia and both the flexural and torsional stiffnesses 
	using equations (\ref{match}) and (\ref{zozza}). In our design of cloaking it is $t=3$ mm, so that it is obtained: $b\simeq5.977$ mm, $\rho\simeq3.05$ g/cm$^3$ and $m_i\simeq2.64$ g.
	Moreover, although $E\simeq12972.3$ MPa was obtained, it was found that the slightly smaller value $E\simeq11610.9$ MPa, calculated considering the reinforcement frame centered at the void edge, performs a little bit better and therefore this value has been assumed in the simulations. 
	Results of f.e.m. (Abaqus) simulations are reported in Fig. \ref{FigPOC_SIM_quantitative} (a) and (b), 
	for the scattering reduction coefficient, equation (\ref{ciccio}), 
	reported for frequencies $f$ ranging between 0 and 700 Hz [and between 0 and 400 Hz for part (c)], the broadest frequency interval ever explored. 
	The scattering reduction coefficient is reported on the left of the figure for parts (a), (b), (c), while on the right a sketch of the three set-ups used for simulations, respectively the intact plate and the plates with cloaked and uncloaked voids, are reported.
	Part (a) refers to the 
	same set-up explored with the experiment, 
	while in part (b) the three voids have random orientations. 
	It is shown that the 
	scattering reduction coefficient, equation (\ref{ciccio}),
	ranges within the intervals 
	(0.0056, 0.5238), (0.0122, 0.5870), and (0.0025, 0.6913) for the cases of part (a) and (b) and (c) of Fig. \ref{FigPOC_SIM_quantitative}, respectively. These values show that 
	the cloaking is excellent, with a
	scattering reduction strongly dependent on the frequency, so that it goes up to the values 99.44\%, 98.78\%, and 99.75\% for certain frequencies in the three cases.
	Part (c) of Fig. \ref{FigPOC_SIM_quantitative} 
	is relative to seven randomly distributed and oriented voids
	and 
	compares results in which: (i.) bending and torsional elastic stiffnesses and mass moment of inertia have been matched with the cloaking (triangle/red markers), 
	(ii) bending and torsional elastic stiffnesses are only matched, so that mass redistribution is neglected (square/green markers), and (iii.) only the mass and its second moment are matched, while the elastic stiffness redistribution is neglected (circle/blue markers). It is evident that the mass redistribution (iii.) is worse than nothing, as all the values are higher than 1 and even touch 4. 
	Case (ii.) is not terribly bad: in many cases 
	the scattering reduction coefficient falls below 1 (showing scattering reduction), while 
	for rare values of frequencies the coefficient becomes even close to zero (for instance, 207.84 Hz and 269.53 Hz), thus showing an excellent cloaking effect.
	However, for several frequencies (for instance,  0.26 Hz and 272.81 Hz) the reinforcement works badly and is worse than nothing.
	For this reason the broadband quality of a cloaking should always give evidence.

	The red markers show that the complete matching of stiffness and mass inertia proposed in the present article works very well, with frequencies in which the reduction coefficient is near to zero (goes down to 0.002485). 
	This is another evidence of the efficiency of our cloaking strategy, which anyway is shown to work well within an extremely broad broad range of frequencies.
	The modes associated with three frequencies for which the cloak works exceptionally well, so that multiple and omnidirectional invisibility is almost perfect)
	are reported in part (d) of Fig.  \ref{FigPOC_SIM_quantitative}, where 
	the plate with cloaked voids is shown to behave almost identically with the intact plate, while it strongly differs from the perforated plate, which remains almost undeformed (because it vibrates far from resonance). 
	Additional numerical simulations, experiments, and a movie are reported in the electronic supplementary material.

	\section{Conclusions: a new design paradigm for invisibility}
	
	An efficient and simple procedure has been introduced for the elimination of the scattered field in a flexural lattice or in an elastic plate containing square voids.
	Two principles have been used in the design of the cloaking: 
	the mass and moment of inertia conservation and a stiffening of the boundary of the void with a reinforcement providing the same bending and torsional stiffness of the plate eliminated with the void. A constructive and simple design method has been proposed and verified through both numerical simulations and experiments (involving multiple interacting voids of different inclinations) in a range of frequencies so broad that it can be pointed out that invisibility
	is almost achieved under all dynamic conditions. 
	The effectiveness of our reinforcing technique is presented with reference to the 7 requirements for cloaking that we suggest should be used as a
	reference for evaluating cloaking performances. 
	However, the proposed design has to be considered an initial idea towards the achievement of an \lq ideal' cloak. In fact, the introduced technique is not unique and lends itself to optimization (in terms of material properties, geometry and mass distribution of the reinforcement) to improve cloaking.
	The movies of experiments and simulations (electronic supporting material) will easily convince the civil engineer that a perforated load-bearing wall can be designed to become {\it dynamically identical} to an unperforated wall. Therefore, during an earthquake, the perforated wall will vibrate the same as its unperforated version, which represents an important realization in earthquake engineering.  
	
	\vspace*{5mm}

\footnotesize{
	\noindent  \textbf{Authors' contributions.} {The idea of reinforcement cloaking was generated and developed by all authors. D.M. programmed the codes and developed all numerical simulations. Experimental set-ups were designed by D.M., with the cooperation of D.B. All experiments, including cutting of the samples with a CNC milling machine and practical implementation of the reinforcement and mass redistribution, their complete instrumentation, recording, and video registration, were realized by D.M. at the \lq Instability Lab' of the University of Trento. All authors contributed to discuss the experimental and numerical data and to write the paper.\\
		
\noindent \textbf{Funding.} D.M. and A.B.M. gratefully acknowledge financial support from the \lq ERC Advanced Grant Instabilities and nonlocal multiscale modelling of materials' FP7-PEOPLE-IDEAS-ERC-2013-AdG (2014-2019). D.B. thanks financial support from PRIN 2015 LYYXA8-006. \\
	
\noindent \textbf{Acknowledgements.} The authors thank Mr. F. Vinante for assistance in the experiments.
}


\end{document}